\documentclass[aps,twocolumn,english,floatfix]{revtex4}
\usepackage{amssymb}
\usepackage{natbib}
\usepackage{graphicx}
\usepackage{amsmath}
\usepackage{wasysym}
\usepackage[bookmarks = false]{hyperref}
\usepackage{color}

\begin{document}

\title{Time-resolved boson sampling with  photons of different colors}

\author{Xu-Jie Wang$^{1,\,2,\,3,\,*}$}
\author{Bo Jing$^{1,\,2,\,3,\,*}$}
\author{Peng-Fei Sun$^{1,\,2,\,3}$}
\author{Chao-Wei Yang$^{1,\,2,\,3}$}
\author{Yong Yu$^{1,\,2,\,3}$}
\author{Vincenzo Tamma$^{4,\,5}$}
\author{Xiao-Hui Bao$^{1,\,2,\,3}$}
\author{Jian-Wei Pan$^{1,\,2,\,3}$}
\affiliation{$^1$Hefei National Laboratory for Physical Sciences at Microscale and Department
of Modern Physics, University of Science and Technology of China, Hefei,
Anhui 230026, China}
\affiliation{$^2$CAS Center for Excellence and Synergetic Innovation Center in Quantum
Information and Quantum Physics, University of Science and Technology of
China, Hefei, Anhui 230026, China}
\affiliation{$^3$CAS-Alibaba Quantum Computing Laboratory, Shanghai 201315, China}
\affiliation{$^4$Faculty of Science, SEES, University of Portsmouth, Portsmouth PO1 3QL, UK}
\affiliation{$^5$Institute of Cosmology \& Gravitation, University of Portsmouth, Portsmouth PO1 3FX, UK}
\affiliation{$^*$These two authors contributed equally to this work.}

\begin{abstract}
  Interference of multiple photons via a linear-optical network has profound applications for quantum foundation, quantum metrology and quantum computation. Particularly, a boson sampling experiment with a moderate number of photons becomes intractable even for the most powerful classical computers, and will lead to ``quantum supremacy". Scaling up from small-scale experiments requires highly indistinguishable single photons, which may be prohibited for many physical systems. Here we experimentally demonstrate a time-resolved version of boson sampling by using photons not overlapping in their frequency spectra from three atomic-ensemble quantum memories. Time-resolved measurement enables us to observe nonclassical multiphoton correlation landscapes. An average fidelity over several interferometer configurations is measured to be 0.936(13), which is mainly limited by high-order events. Symmetries in the landscapes are identified to reflect symmetries of the optical network. Our work thus provides a route towards quantum supremacy with distinguishable photons.
\end{abstract}

\maketitle

Universal linear-optical quantum computing~\cite{Kok2007a} is generally considered to be challenging in the near future. An intermediate quantum computing model, namely ``boson sampling" which requires less demanding experimental overheads~\cite{aaronson2011computational}, has attracted intensive experimental interests in recent years ~\cite{spring2013boson,broome2013photonic,tillmann2013experimental,Crespi2013c,Spagnolo2014,Carolan2014,Carolan2015,Bentivegna2015,wang2017high}. A boson sampling machine can be realized by interfering many single photons through a linear optical network. Sampling the output photon distribution is strongly believed to be intractable for a classical computer for large photon numbers~\cite{aaronson2011computational,Rohde2012}. For experimental realizations, photon indistinguishability is crucially important, since for distinguishable photons the computational complexity collapses to a polynomial scaling, which becomes tractable for a classical computer~\cite{Rohde2012}. Requiring of complete overlap in the photonic spectra as a way to achieve photon indistinguishability may impose a challenge for many types of photon sources, particularly the solid-state single photon emitters~\cite{Aharonovich2016}. The inhomogeneous distribution of complex mesoscopic environment of the solid state tend to cause frequency distinguishability for photons created from different emitters.

Photon indistinguishability and interference are also very important fundamentally. The HOM dip  is a beautiful manifestation of the interference of two identical photons~\cite{alley1986proceedings,Hong1987,shih1988new,Walmsley2017}. When two photons are different in color, it was shown that high interference visibility can be recovered by using time-resolved measurement~\cite{Legero2003b,Legero2004}. Perfect coalescence can still happen when photons are detected simultaneously in the two output modes. Later it was studied that perfect entanglement swapping by interfering color-different photons is also possible by using time-resolved measurement and active feed forward~\cite{Zhao2014}. Very recently, Tamma and Laibacher showed that by using polarization- and time-resolved detections at the output of a random linear optical network, a much richer multiphoton correlation landscape can be observed for a Boson sampling experiment with photons not overlapping in their temporal or frequency spectra and photons with  different polarizations~\cite{tamma2015multiboson}. They also proved that the computational complexity is at least as hard as in standard boson sampling~\cite{laibacher2015physics,tamma2016multi,laibacher2018toward}.

In this paper, we report for the first time a time-resolved boson sampling experiment when no overlap occurs between the photonic spectra\cite{tamma2015multiboson,laibacher2015physics}. We make use of three cold atomic ensembles to create three independent single photons, which are injected into a linear optical network with its internal phase being adjustable. At the output ports of the network, the three photons are detected in a time-resolved manner. We observe different kinds of multiphoton correlation landscapes as we change the phase configuration. The observed coincidence landscapes agree very well with theoretical calculations. Moreover, we also find that symmetries in the multiphoton coincidence landscape can reveal symmetries of the optical network~\cite{laibacher2017spectrally}. Our work enables multiphton interference and distinguishability to be recovered by using time-resolved measurements, and thus provides a route towards demonstrating quantum supremacy~\cite{Preskill2012,Harrow2017,laibacher2018toward} with nonidentical photons.

\begin{figure}[htbp]
\includegraphics[width=\columnwidth]{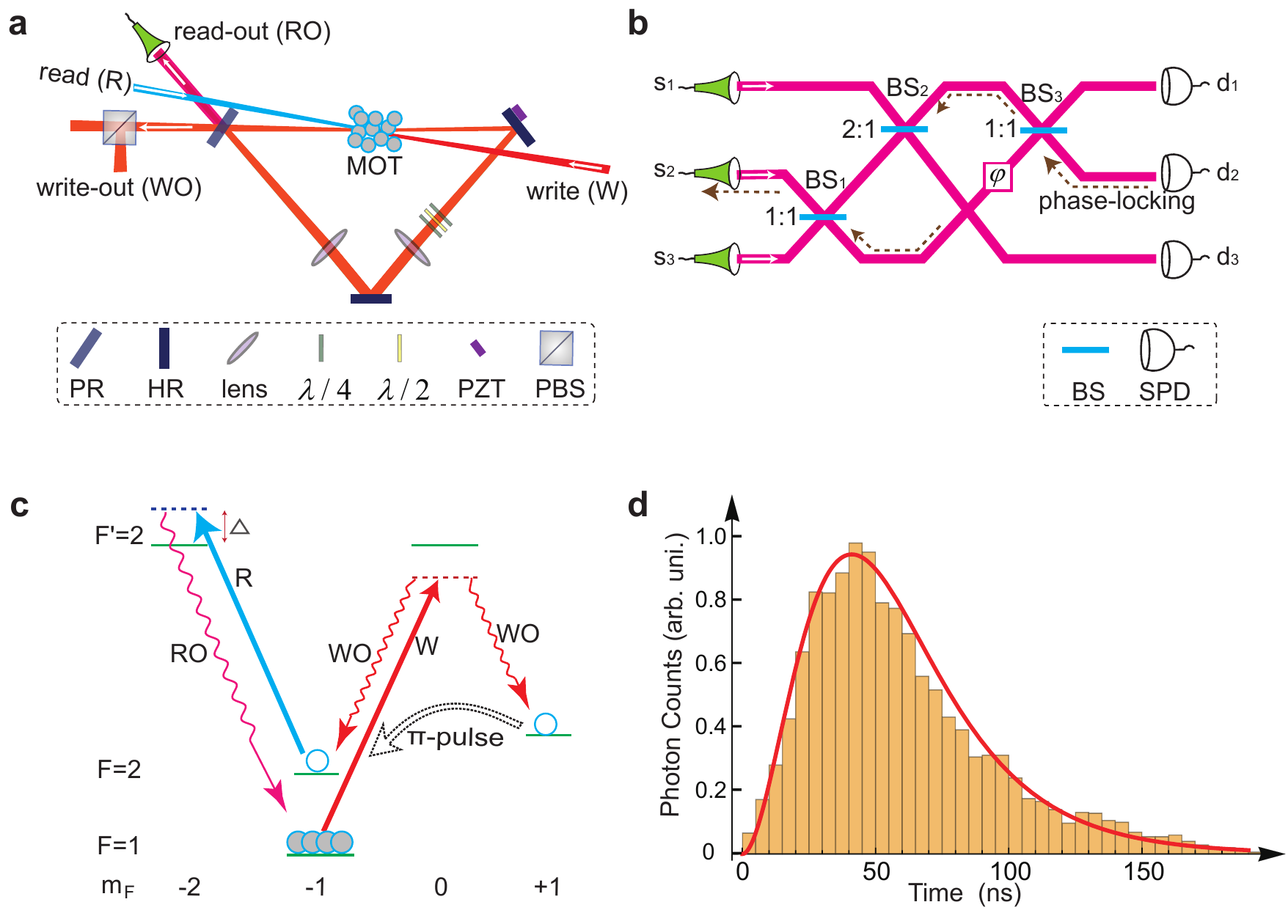}
\caption{\textbf{Experimental setup.~}\textbf{a}, Schematic diagram for one setup of atomic ensemble quantum memory to create one single photon. A ring cavity, which mainly consists of one partially reflecting mirror (PR, $R\simeq 80\%$) and two highly reflecting  mirrors (HR, $R\ge 99.9\%$), is used to enhance single photon generation rate. The half- and quarter-wave plates ($\lambda /2$ and $\lambda /4$) in the cavity are used for polarization compensation. The cavity is intermittently locked by a piezoelectric ceramic transducer (PZT). In our boson sampling experiment, we make use of three similar setups to create three single photons. \textbf{b}, Linear optic network to implement boson sampling. The network is mainly composed of three beam splitters (BS, $R:T=1:1$ or $2:1$). Heralded single photon efficiency from atoms to the network is 45\%.  \textbf{c}, Atomic levels used in the single-photon source.  Atoms are prepared at state  $|1,-1\rangle_{F,mF}$ by optical pumping during each write process. After applying the write pulse, there are two orthogonal polarized write-out photons collected, once the spinwave is produced in state $|2,+1\rangle_{F,mF}$, a $\pi$-pulse is applied to transfer the excitation to state $|2,-1\rangle_{F,mF}$. \textbf{d} A typical read-out single-photon profile. The histogram is a measured photon counts distribution by a single-photon detector during coincidence measurement. The solid line represent the read-out photon profile that we used in theoretical calculation.}
\label{fig:Setup}
\end{figure}

In our experiment we make use of a versatile setup of cold atoms~\cite{yang2016efficient} to create single photons~\cite{Chen2006,Matsukevich2006b}. An atomic ensemble is captured and cooled through magneto-optical trapping (MOT). By employing the spontaneous Raman scattering process with a $\Lambda$ energy scheme, we can create nonclassical correlations between a scattered photon and a spinwave excitation in a probabilistic way~\cite{Duan2001}. The spinwave excitation can be later retrieved as a second photon on demand.  The experimental scheme is depicted in Fig.~\ref{fig:Setup}.  To suppress high-order events, the excitation probability in the write process is typically very low. In order to enhance the photon generation rate, we make use of a dual Raman scattering process. A $\sigma^+$ polarized write-out photon heralds a $|2,-1\rangle_{F,mF}$ collective excitation. While a $\sigma^-$ polarized write-out photon heralds a $|2,+1\rangle_{F,mF}$ collective excitation, and we conditionally apply a $\pi$ pulse which transfer the $|2,+1\rangle_{F,mF}$ excitation to $|2,-1\rangle_{F,mF}$ excitation. The $|2,-1\rangle_{F,mF}$ excitation is later retrieved as a single photon on demand. Polarization multiplexing enables us to double the photon generation rate without increasing the contribution of high-order events.

\begin{table}[htbp]
	\caption{Measurement of the second-order auto-correlation $g^{(2)}$.}
	\label{table1}
	\begin{ruledtabular}
		\begin{tabular}{c@{}c@{}c@{}c@{}c@{}c@{}c@{}}
			
			$p_e$	&0.01  &0.02  & 0.03  & 0.04   &
			0.06    \\ \hline
			$g^{(2)}(s_1)$	&0.072(7)  &0.126(9)  & 0.201(11)  & 0.233(11)   & 0.322(12)    \\			
			$g^{(2)}(s_2)$	&0.094(9)  &0.165(10)  & 0.222(11)  & 0.279(11)   & 0.335(13)    \\
			$g^{(2)}(s_3)$	&0.110(9)  &0.142(8)  & 0.220(101)  & 0.251(10)   & 0.361(13)    \\
		\end{tabular}
	\end{ruledtabular}
\end{table}

To demonstrate multiphoton boson sampling, we make use of three similar setups to create three single photons as shown in Fig.~\ref{fig:Setup}. We first measure the single photon qualities. For each setup, we repeat the write process until a write-out photon is detected, and retrieve the heraldedly prepared spin wave excitation afterwards to a single photon. We measure the second-order auto-correlation $g^{(2)}$ for the retrieved photon as a function of retrieval time. Within a storage duration of 50 $\mu$s, we find that the parameter $g^{(2)}$ hardly changes for each setup. We also change the excitation probability $p_e$ and measure the parameter $g^{(2)}$ accordingly for each setup, with the results shown in Table~\ref{table1}. Each value is averaged over the range of $0\sim50$ $\mu$s. The $g^{(2)}$ parameter under the same write-out probability is nearly the same for three setups. In our experiment, we set the excitation probability to be $p_e=0.04$ by making compromise between single-photon generating rate and the single-photon quality.

It is crucial for a boson sampling experiment that many single photons are released simultaneously.  For traditional photon sources like spontaneous parametric down-conversion or spontaneous four wave mixing, this requirement imposes a scalability issue, since heralded photons are generated randomly in time and simultaneous creation is rare. An additional quantum memory may be employed to solve this issue~\cite{Kaneda2015,Kaneda2017}. More recently, the so-called scattershot multiboson correlation sampling problem was introduced by allowing additional sampling in the photonic inner degrees of freedom (e.g. central frequencies and times) at the interferometer input and output\cite{laibacher2018toward}. In this case, the classical hardness of approximate boson sampling experiments emerges also for photons emitted at random time or with random colors\cite{laibacher2018toward}. In our experiment however, the on-demand character of the photon source directly enables us to create multiple photons in an efficient way. If the memory lifetime is long enough, we can simply repeat the write process for setup until success and simultaneously retrieve three photons when all setups are ready. While our current setup has a limited lifetime ($\sim 64 {\rm \mu s}$), thus we set a maximal trial number of $m=7$. If less than three setups are ready when maximal trial limit is met, we restart the preparation process, otherwise we retrieve the three photons simultaneously. Such a preparation process enhances the $n$ photon rate by a factor of $\left[ 1-(1-p_e)^m \right]^n/(mp_e^n)$.

\begin{figure*}[htbp]
\includegraphics[width=1.3\columnwidth]{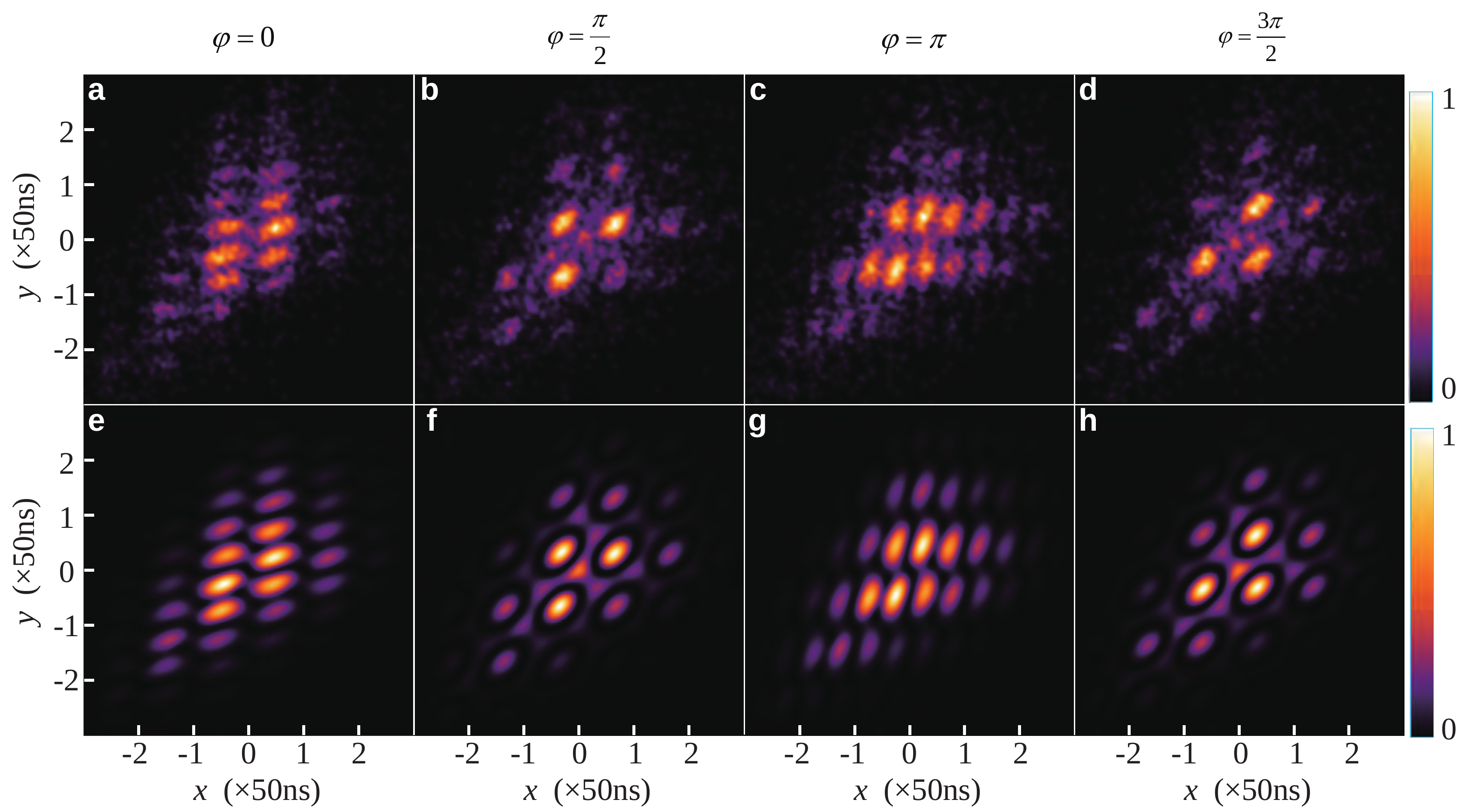}
\caption{\textbf{Temporal correlation landscapes under different phase configurations.}
Plots in the first row are experimental results (\textbf{a$\sim$d}). The horizontal axis $x$ in each plot refers to $(t_1 -t_3)$, while the vertical axis $y$ refers to $(t_2 -t_3)$. Each point corresponds to a collection of events centered around $(x,y)$ with a weighting function in the form of $e^{-\left[ (x-x_i)^2+(y-y_i)^2 \right]/r_0^2}$, where $(x_i,y_i)$ refers to the coordinate of a nearby three fold event and the parameter $r_0$ is chosen to be 3 ns to make compromise between fluctuation of event counts and reduction of pattern resolution. Plots in the second row are theoretical results (\textbf{e$\sim$h}). Fidelities between experimental and theoretical results are calculated to be  $F(0)= 0.945$, $F(\pi /2)= 0.928$, $F(\pi)= 0.950$, $F(3 \pi /2)= 0.923$ respectively.}
\label{fig2}
\end{figure*}

The prepared multiple single photons are coupled into a multiport interferometer, which is constructed using bulk linear optics as shown in Fig.~\ref{fig:Setup}b. The internal phase is actively stabilized to an adjustable value $\varphi$. Photons at each output mode are detected with a single-photon detector (SPD). All detected events are registered with a multichannel time-to-digit converter (TDC), and from which we can analyse multifold temporal correlations. To demonstrate boson sampling with color-different photons, the three photons are blue detuned by $2\pi \times 72.4$ MHz ($s_1$), $2\pi \times 33.0$ MHz ($s_2$), $2\pi \times 52.4$ MHz ($s_3$) relative to the D1-line transition $\left|F\!=\!1\right>$ $\leftrightarrow$ $\left|F'\!=\!2\right>$ by adjusting the read beam frequency accordingly for each source. We make measurements for a number of different phases $\varphi$, and analyze multiphoton temporal correlations, with results shown in Fig.~\ref{fig2}a-d.

\begin{figure}[htbp]
\includegraphics[width=\columnwidth]{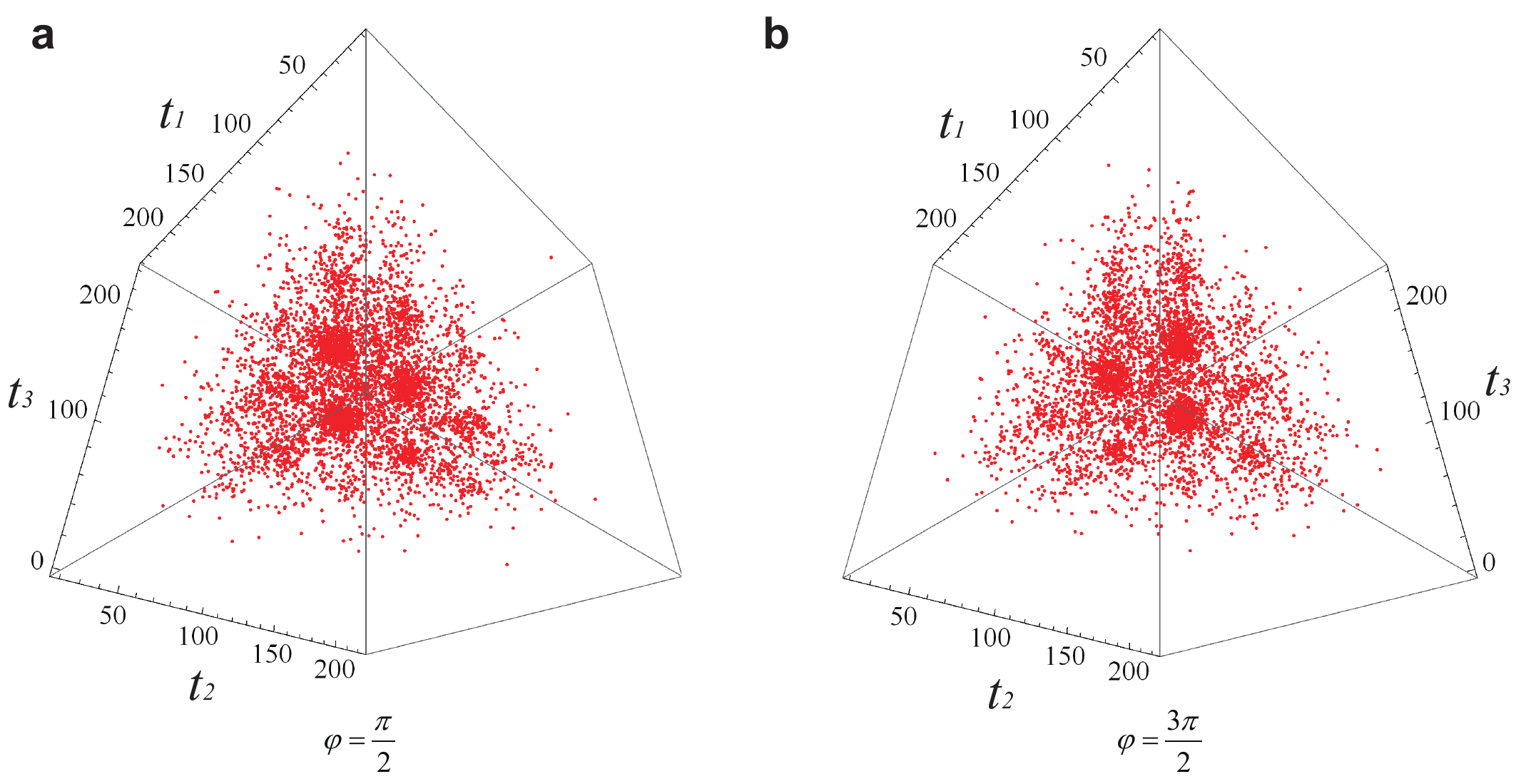}
\caption{\textbf{Measured temporal correlation landscapes in a 3-dimension coordinate.} \textbf{a}, $\varphi$ is set to $\pi /2$. \textbf{b}, $\varphi$ is set to $3\pi /2$. Both plots are viewed along the direction $(1, 1, 1)$. Each data point represents a three-photon coincidence event registered at $(t_1,t_2,t_3)$.}
\label{fig3}
\end{figure}

We find that the correlation landscapes have very interesting structures, and changes remarkably as we change the phase $\varphi$. In the case of $\varphi=0$, the unitary transformation is
\begin{equation}
U_0=\frac{1}{\sqrt{3}}
\begin{bmatrix}
    -1 & (\sqrt{3}-1)/2 & i(\sqrt{3}+1)/2 \\
    i & i(\sqrt{3}+1)/2 & (1-\sqrt{3})/2 \\
    1 & -1 & i
\end{bmatrix}.
\end{equation}
The permanent of $U_0$ is 0, which means that interference occurs destructively for all multiphoton paths if the input photons are perfectly identical each other. Thus, no three-fold coincidence will be detected at the output modes. In our experiment however the input photons have different frequencies ($\Delta\omega \geq \delta\omega$, $\delta\omega = 2\pi \times 12.9 \,\rm{MHz}$ is the single-photon linewidth). Nevertheless, the adoption of fast detection enable us to erase the color information, and destructive interference can be recovered if the three photons are detected simultaneously\cite{tamma2015multiboson}. This is clearly proved by Fig.~\ref{fig2}a, as the region around $(0,0)$ is rather dim in comparison with the peaks nearby. Departing from the dip, we observe beating patterns both in the direction of $x \equiv t_1-t_3$ and $y \equiv t_2-t_3$. Based on our Fourier analysis, beating in the x direction has a period of $51.4(11)$ ns, which is mainly due to interference of photon $s_1$ and and photon $s_3$. While beating in the y direction has a period of $24.9(3)$ ns, which is mainly due to interference of photon $s_1$ and and photon $s_2$. In the case of $\varphi=\pi/2$, the interferometer acts as a symmetric tritter which is described as $U_{ds}=e^{i2\pi ds/3}/\sqrt{3}$, where $s=1,2,3$ refers to the modes at the source side and $d=1,2,3$ refers to the modes at the detector side. The multiphoton correlation landscape has a quite different structure, as shown in Fig.~\ref{fig2}b\cite{laibacher2017spectrally}. At the center point, it corresponds to the case of identical photons, for which the output of this tritter is either one photon in each mode or three photons in a same mode~\cite{Spagnolo2013}. Away from the center point, the interference period is measured to be $49.7(7)$ ns and $50.3(13)$ ns respectively for the direction $x$ and $y$, which are mainly due to equal contribution of all three pairwise interferences of the three input photons\cite{laibacher2017spectrally}.

We also find that the observed correlation landscapes have some symmetries which may reflect symmetries either of the photons or the linear optic network~\cite{laibacher2017spectrally}. When $\varphi$ is switched from $0$ to $\pi$ or from $\pi/2$ to $3\pi/2$, it is equivalent of interchanging the mode label 1 and 2 after $BS_3$. Therefore, the correlation landscape of $\varphi=\pi$ ($3\pi/2$) should looks the same as the landscape of $\varphi=0$ ($\pi/2$) if we interchange the axis $x$ and $y$. This is clearly proved by our result shown in Fig.~\ref{fig2}c and Fig.~\ref{fig2}d, comparing with Fig.~\ref{fig2}a and Fig.~\ref{fig2}b respectively. Moreover, in the case of $\varphi=\pi /2$ or $3\pi /2$ the network acts as a symmetric tritter, which should give rise to a three fold symmetry in the correlation landscapes~\cite{laibacher2017spectrally}. Therefore, we replot our experimental data in a 3-dimensional coordinate as shown in Fig.~\ref{fig3}. Each data point represents a threefold coincidence event at the time coordinate ($t_1$, $t_2$, $t_3$). We can clearly identify a threefold rotational symmetry around the axis $(1,1,1)$, and three mirror symmetric planes\cite{laibacher2017spectrally}.

To evaluate how well our experiment demonstrates the boson sampling process, we calculate the corresponding theoretical landscapes and made comparisons. By modeling the temporal wave packet with a function shown in Fig.~\ref{fig2}d, the calculated theoretical landscapes are shown in Fig.~\ref{fig2}e-h. Apparently the experimental and theoretical landscapes resemble each other very well. To make a quantitative evaluation, we define a  fidelity function as~\cite{tillmann2013experimental}
\begin{equation}
F=\frac{\sum\nolimits_{i,j}^{N,M} f_{e}\left( x_i,y_j \right)^{1/2} \centerdot f_{t}\left( x_i,y_j \right)^{1/2}}{\left[  \sum\nolimits_{i,j}^{N,M} f_{e}\left( x_i,y_j \right) \right] ^{1/2} \centerdot \left[\sum\nolimits_{i,j}^{N,M} f_{t}\left( x_i,y_j \right) \right] ^{1/2}},
\end{equation}
where $f_{e}(x,y)$  is the experimental measured probability distribution function, and $f_{t}(x,y)$ is the theoretical distribution. The fidelity gets its maximal value of 1 for two same landscapes. The calculated fidelities for different phases are shown as insets in Fig.~\ref{fig2}, and give an average value of $\overline{F} = 0.936(13)$. We attribute the imperfect fidelity mainly to limited single photon qualities.

In summary, we have demonstrated a time-resolved version of boson sampling with color-different photons. The observed correlation landscapes have very rich structures and shows some form of symmetry which is inherently related with symmetries of the linear-optic network. Besides, the adoption of memory-based photon sources enables efficient creation of multiphoton state via feedback. Moreover, the method of using color-different photons for boson sampling mitigates the requirement in generating identical photons significantly for many physical systems. By employing a deterministic approach of photon creation and efficient coupling, scalable extending the current experiment to more disparate photons will become possible in the near future, and may lead to quantum supremacy with photons in a conceptually new way. This work also motivates future demonstration of the computational hardness of boson sampling with input photons with random colors\cite{laibacher2018toward} as well as novel schemes for the characterization of the evolution of arbitrary single photon states in linear optical networks\cite{tamma2015multiboson,laibacher2017spectrally,zimmermann2017role}.

This work was supported by National Key R\&D Program of China (No. 2017YFA0303902), National Natural Science Foundation of China, and the Chinese Academy of Sciences. V.T. acknowledges partial support from the Army Research Laboratory under Cooperative Agreement Number W911NF-17-2-0179. V.T. also acknowledges useful discussions with Simon Laibacher.

\bibliography{myref}

\end{document}